\documentclass[reprint,aps,longbibliography]{revtex4-1}
\usepackage[utf8]{inputenc}

\usepackage{graphicx}
\usepackage{natbib}
\usepackage{hyperref}
\usepackage{amsmath}
\usepackage{amssymb,mathtools}

\newcommand{\ket}[1]{\ensuremath{\left| #1 \right\rangle}}
\newcommand{\braket}[2]{\ensuremath{\left\langle #1 | #2 \right\rangle}}
\newcommand{\hilb}{\mathcal{H}}
\newcommand{\exch}[1]{\ensuremath{\mathrm{EX}_{#1}}}
\newcommand{\swap}{\ensuremath{\mathrm{SWAP}}}
\newcommand{\sorter}[1]{\ensuremath{S_{#1}}}
\newcommand{\suptext}[2]{\ensuremath{#1^{(\mathrm{#2})}}}
\newcommand{\xgate}{\ensuremath{{X}}}
\newcommand{\xk}{\ensuremath{{X}^k}}

\begin{document}

\title{Arbitrary unitaries in orbital angular momentum of single photons}
\author{Jaroslav Kysela}
\affiliation{University of Vienna, Faculty of Physics, Boltzmanngasse 5, 1090 Vienna, Austria}
\affiliation{Institute for Quantum Optics and Quantum Information, Austrian Academy of Sciences, Boltzmanngasse 3, 1090 Vienna, Austria}

\begin{abstract}
A simple argument is presented that explicitly shows how to construct an arbitrary quantum gate acting on orbital angular momentum (OAM) of single photons. The scheme can be applied to implement subspace multiplexing, where a single high-dimensional OAM qudit represents effectively a stack of multiple independent lower-dimensional qudits. A special subclass of unitaries composed of single-photon controlled gates is studied in detail and notable examples of the general approach are discussed. The generalization of the simple argument leads to the parallelization scheme, which results in the savings of resources. The presented schemes utilize only conventional optical elements and apply not only to single photons but also to classical light.
\end{abstract}

\maketitle

\section{Introduction}

The orbital angular momentum (OAM) of a photon amounts to quantized twists of the photon's wave function \cite{erhard_twisted_2018,shen_optical_2019} and has served in a multitude of experiments as a high-dimensional quantum carrier of information---for given $d$ one can always consider a $d$-dimensional subspace of OAM spanned by eigenstates $\ket{0}, \ldots, \ket{d-1}$, where the information is encoded in a superposition of these eigenstates. The OAM of photons has been experimentally utilized in quantum teleportation \cite{Wang2015}, high-dimensional quantum key distribution \cite{Bouchard2018experimental}, generation of high-dimensionally entangled quantum states \cite{dada_experimental_2011,PhysRevA.86.012334}, as well as in more fundamental experiments that study the correspondence principle for a very high number of OAM quanta \cite{Fickler13642}.

To make OAM of photons a full-fledged degree of freedom suitable for information transmission and processing, it is necessary to be able to manipulate the information contained in the OAM states as required by a given application. Mathematically speaking, one should be able to apply an arbitrary unitary operation to the quantum state of OAM. In this paper, we present a very simple argument that demonstrates that any unitary can be implemented using only conventional optical elements. We also introduce a scheme that allows to construct a single-photon controlled-gates, where either the control or the target qudits are played by the path degree of freedom of the photon. The direct generalization of these schemes leads one to the study of the parallelization scheme where a series of simultaneously applied identical local unitaries is replaced with only a single instance of the unitary supplemented with pre- and post-processing stages.

The schemes studied in the text amount to networks of interferometers. In real experimental conditions, the stability of interferometers is in general an important issue that hinders further development of this technology and over the years reliable holographic techniques have been developed \cite{Morizur10} and successfully demonstrated \cite{Labroille2014,Fontaine2019,Brandt20} that overcome the stability problem. Being aware of that, the aim of the present paper is not just to present an alternative point of view, based on interferometers, but also to emphasize some of the algebraic properties related to the orbital angular momentum. One such property of the resulting setups is periodicity in OAM, which can in principle be used for multiplexing multiple OAM quantum states lying in different subspaces into a single large superposition.

The OAM eigenstates are represented by wavefunctions whose form in cylindrical coordinates $(r, \phi, z)$ takes on the form
\begin{equation}
    \phi_k(r, \phi, z) = \braket{r, \phi, z}{k} \propto e^{i k \phi},
\end{equation}
where $k$ is an integer. This wavefunction undergoes simple transformations when the corresponding photon is subjected to the action of various conventional optical elements. Throughout the paper, we consider only the following toolkit of optical elements: mirrors, beam splitters, Dove prisms, phase shifters, and simple holograms in the form of spiral phase plates. Most notably, no complicated phase profiles resulting from numerical simulations are used that would require the use of costly spatial light modulators. The summary of actions of the individual elements on OAM states can be found in Ref.~\cite{oamfft2}.

The manuscript is structured as follows. At first, we demonstrate in Sec.~\ref{sec:universal} that conventional optical elements are sufficient to construct an arbitrary unitary gate in OAM of single photons. To exemplify this approach, we present an implementation of high-dimensional Pauli gates and their integer powers. Such gates are used for example in the construction of Heisenberg-Weyl observables, which in turn find applications in the quantum state tomography \cite{HeisenbergWeylObs,Palici_2020}. After that we turn our attention to the single-photon controlled-gates in Sec.~\ref{sec:controlled}, where the OAM plays the role of the control qudit and the path degree of freedom of the same photon represents the target qudit. The opposite case with the two roles exchanged then follows from the latter by using a swap operator. We conclude the list of interferometric networks in Sec.~\ref{sec:parpaths}, where the parallelization scheme is studied. It allows to replace a series of identical local unitaries by a single setup that contains significantly reduced number of optical elements. To give a specific example, we present explicit setups for parallelized Pauli gates. The general schemes can be simplified using the polarization of photons. Also, the schemes exhibit periodicity that is discussed in Sec.~\ref{sec:periodicity}. We summarize our results in Sec.~\ref{sec:conclusion}.

\section{Arbitrary unitaries in OAM}
\label{sec:universal}

\subsection{General case}

One of the core results of the quantum computation theory is that there exist universal sets of operations, out of which any other unitary operation can be constructed. For the case of operations acting on the OAM of a single photon, such universal sets have been presented in Refs.~\cite{Garcia-Escartin2011,Gao2019}. Here we put forward a very simple argument that shows explicitly that such a universal set can be constructed only from conventional optical elements.

Let us denote by $U$ the abstract $d$-dimensional unitary operation and let $U_O$ and $U_P$ be its implementations for the OAM and the path degrees of freedom, respectively. The idea underlying our argument is this: to build $U_O$ one first transforms the incoming OAM eigenstates into the path encoding and then applies $U_P$, for which general implementation schemes are known \cite{Reck1994,Clements2016,Guise2018} that use only beam splitters and phase shifters. At the end, the propagation modes are transformed back into the OAM eigenstates. The transition between the OAM and path encodings is performed by a $d$-dimensional OAM sorter $\sorter{}$. The sorter turns an OAM eigenstate $\ket{m}_O$ with $m$ quanta of OAM and propagating along the zeroth path $\ket{0}_P$ into the fundamental mode $\ket{0}_O$ that propagates along the $m$-th path $\ket{m}_P$, such that
\begin{equation}
    \sorter{}(\ket{m}_O\ket{0}_P) = \ket{0}_O\ket{m}_P.
    \label{eq:sorter}
\end{equation}
The OAM sorter can be implemented in multiple ways \cite{Berkhout2010,Labroille2014,Fickler2017,Fontaine2019}. Here we employ the interferometric implementation that consists merely of beam splitters, Dove prisms, and holograms \cite{Leach2002,Leach2004,Garcia-Escartin2008,oamfft} and whose structure is shown in section \ref{sec:pauli}. The whole scheme is then compactly represented by the formula
\begin{equation}
    U_O = \sorter{}^{-1} \cdot U_P \cdot \sorter{}.
    \label{eq:universal}
\end{equation}
In the scheme, the use is made of $O(d^2)$ beam splitters, $O(d^2)$ phase shifters, $O(d)$ Dove prisms, and $O(d)$ holograms as can be deduced from the structure of the OAM sorter \cite{xgate,oamfft} and the Reck \emph{et al.} scheme \cite{Reck1994}. The detailed analysis of the number of optical elements can be found in Appendix \ref{sec:num_of_elems} and the effect of losses is briefly studied in Appendix \ref{sec:losses}.

Despite the ``obviousness'' of the universal scheme in Eq.~\eqref{eq:universal}, the author of these lines has not found any publication so far that would mention it. The scheme offers some advantages when compared to alternative approaches. It works in a deterministic and in principle lossless way, unlike the approach based on the decomposition of a general unitary into a linear combination of powers of $\xgate$ and $Z$ gates \cite{erhard_twisted_2018}. There is also no need to assume a limit of many elements as in Ref.~\cite{Garcia-Escartin2011}, no complicated theoretical framework is necessary to demonstrate universality as in Ref.~\cite{Gao2019}, and there is also no need to perform numerical Fourier-optics optimization algorithms as in Ref.~\cite{Morizur10}.

\subsection{High-dimensional Pauli gates}
\label{sec:pauli}

\begin{figure*}
\includegraphics[width=0.9 \linewidth]{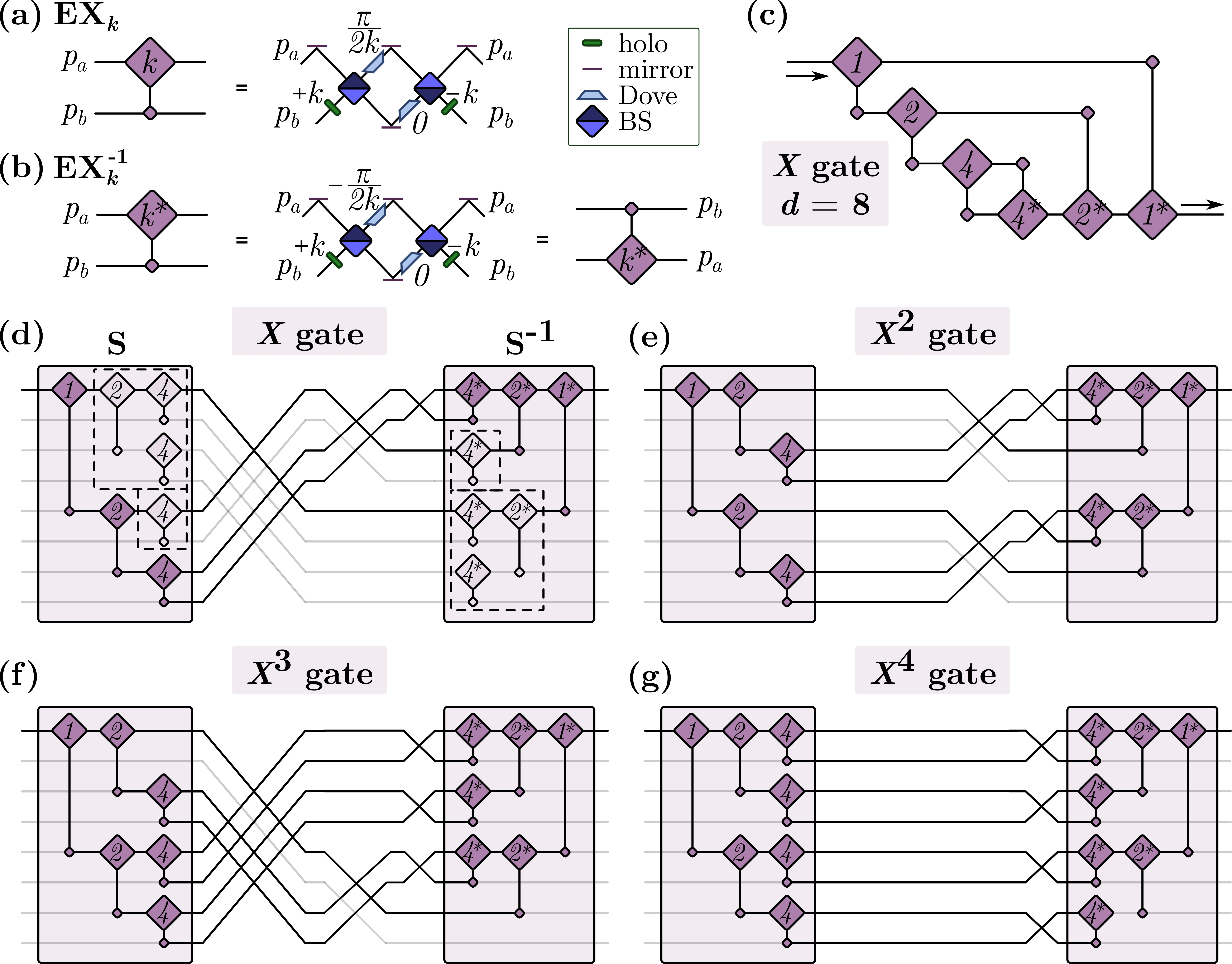}
\caption{The interferometric implementation of the $\xgate$~gate together with its integer powers. 
(a) The OAM exchanger $\exch{k}$ of order $k$ is built from two holograms and a Leach interferometer \cite{Leach2002} with one Dove prism rotated through $\pi/(2 k)$. Optical elements: holo---hologram, Dove---Dove prism, BS---50:50 beam splitter. 
(b) The inverse of the OAM exchanger, $\exch{k}^{-1}$, has almost the identical structure to that of $\exch{k}$ only the Dove prism is rotated through $-\pi/(2 k)$. For convenience, we use two slightly different symbols to denote the inverse of the OAM exchanger, as shown in the figure. 
(c) The $\xgate$~gate for $d = 8$. 
(d) The $\xgate$~gate from (c) is constructed from two OAM sorters, marked by shaded rectangles in the figure, from which redundant exchangers are removed. These exchangers can be grouped into blocks of increasing size, which are enclosed in dashed-line rectangles. The unused paths as well as redundant exchangers are drawn in faded color. The remaining exchangers can be reordered in order to get rid of the path permutations. 
(e) The same principles apply when constructing the integer powers $\xk$ of the $\xgate$~gate. When the exponent $k$ is a power of two, i.e., $k = 2^m$, the path permutation has a repetitive structure and the whole setup is effectively split into $k$ identical subsetups. 
(f) For a general exponent $k$ the path permutation has a more complicated structure. 
(g) The number of exchangers that have to be retained in the final setup increases with the exponent $k$ until it attains the form $k = d/2$, in which case no exchangers can be removed.}
\label{fig:xgate}
\end{figure*}

The brute-force scheme of Eq.~\eqref{eq:universal} can be simplified considerably for specific unitary operations. One class of such operations are the Pauli operators, prominent examples of local quantum gates. The $d$-dimensional Pauli $\xgate$~gate and $Z$~gate are defined by \cite{HeisenbergWeylObs}
\begin{eqnarray}
    \xgate_d (\ket{q}) & = & \ket{(q + 1) \ \mathrm{mod} \ d}, \label{eq:xgate} \\
    Z_d (\ket{q}) & = & \omega^{q} \ket{q}, \label{eq:zgate}
\end{eqnarray}
where $\omega = \exp(2 \pi i/d)$, and where $\{ \ket{q} \}_{q=0}^{d-1}$ form the computational basis. It turns out that to implement the $Z$ gate as well as its integer powers $Z^k$ a single optical element is sufficient---a Dove prism rotated through an angle of $k \pi/d$ performs the required transformation. The approach of Eq.~\eqref{eq:universal} is thus not necessary in this case. However, for the $X$ gate the situation is more complicated and the detailed analysis, based on utilizing Eq.~\eqref{eq:universal}, is presented below.

The implementation scheme of the $d$-dimensional $\xgate$~gate acting on OAM was given in Refs.~\cite{xgate,isdraila_cyclic_2019}. In what follows, we generalize results of Ref.~\cite{xgate} and construct a setup for a $k$-th power of the $\xgate$~gate, $\xk$~gate, in a way that is more efficient than a mere concatenation of $k$ setups corresponding to the $\xgate$~gate. It turns out that the most resource-demanding case is that of $k = d/2$, for which the scaling of resources is linear $O(d)$. An alternative approach in Ref.~\cite{Palici_2020} for constructing $\xk$~gates scales like $O(d \log_2(d))$.

We make use of OAM exchangers, depicted in Fig.~\ref{fig:xgate}(a), which are passive two-input two-output optical devices \cite{oamfft} composed of a Leach interferometer \cite{Leach2002} and two holograms. The sorting properties of an OAM exchanger $\exch{k}$ of order $k$ are determined by the value of $k$ and so is the case for the inverse operation $\exch{k}^{-1}$ shown in Fig.~\ref{fig:xgate}(b). The $\xgate$~gate can be built out of OAM exchangers in an arbitrary dimension. Nonetheless, here we focus only on dimensions of the form $d = 2^M$, for which the $\xgate$~gate can be constructed as a series of OAM exchangers of orders $2^k$ for $k = 0, \ldots, M-1$, followed by the reversed series of the same structure. The modification of the original scheme for dimension $d = 8$ is presented in Fig.~\ref{fig:xgate}(c) and the generalizations for higher dimensions follow analogously the depicted pattern. This pattern can be obtained by starting from the naive implementation in Eq.~\eqref{eq:universal}, where the OAM sorters are constructed as binary-tree networks of OAM exchangers \cite{Leach2002,oamfft}. This case is explicitly shown in Fig.~\ref{fig:xgate}(d), where the path-encoded implementation $U_P$ of the $\xgate$~gate corresponds to the path permutation that connects the output ports of the sorter $\sorter{}$ on the left with the input ports of the inverted sorter $\sorter{}^{-1}$ on the right. Due to the structure of the path permutation, many exchangers $\exch{k}$ from $\sorter{}$ are followed by their inverses $\exch{k}^{-1}$ from $\sorter{}^{-1}$. All these exchangers can be obviously removed without any effect on the final state. The resulting optical network is identical to that in Fig.~\ref{fig:xgate}(c).

The $\xgate$~gate is a specific example of a cyclic permutation of the basis states. We can obtain all other cyclic permutations by taking powers of the $\xgate$~gate. Specifically
\begin{equation}
    \xgate_d^k (\ket{q}) = \ket{(q + k) \ \mathrm{mod} \ d},
\end{equation}
where $k \in \mathbb{N}$. Due to the cyclic property of the $\xgate$~gate, it holds that $\xgate^{d-k} = (\xgate^k)^{-1}$. Consequently, it suffices to study only powers $k \le d/2$ as the implementation of $\xk$ for $k > d/2$ is obtained as the implementation for $\xgate^{d-k}$ operated backwards. We proceed analogously to the case of the $\xgate$~gate. We again start from the general scheme in Eq.~\eqref{eq:universal} and remove all the exchangers that do not affect the final state. In Figs.~\ref{fig:xgate}(e), (f), and (g) the explicit form of $\xk$~gate is shown for $d = 8$ and $k = 2, 3, 4$, respectively.

In order to understand how the scheme in Eq.~\eqref{eq:universal} can be simplified for general dimension $d = 2^M$ and general power $1 \le k \le d/2$ one notes that the permutations of propagation paths can be expressed as a series of path crossings of increasing size\footnote{The reason why the path permutations themselves in Fig.~\ref{fig:xgate} are not cyclic permutations is that there is an additional permutation that comes from the construction of the OAM sorter $S$ and its inverse $S^{-1}$ \cite{oamfft}.}, see the middle part of Fig.~\ref{fig:xgate}(d). When $k$ is a power of two, i.e., $k = 2^m$, the path permutation has a repetitive structure, cf. Fig.~\ref{fig:xgate}(e) and (g). In such cases, the setup effectively decomposes into $k$ subsetups of the same structure and smaller size. For example, the setup of $\xgate^2$ in Fig.~\ref{fig:xgate}(e) in dimension $d = 8$ can be viewed as two smaller setups for $\xgate$ in dimension $d = 4$. In general, the setup of $\xk$ for power $k = 2^m$ in dimension $d = 2^M$ can be seen as $k$ subsetups implementing $\xgate$~gate in dimension $d' = d/k = 2^{M-m}$. These setups are sandwiched between two OAM sorters with $k$ output paths. This way we obtain the simplified scheme for general dimensions $d$ and powers $k$ that are both powers of two. To characterize the general structure of the setup for $\xk$ when the power $k$ is not a power of two, such as the case of $\xgate^3$ in Fig.~\ref{fig:xgate}(f), is more subtle. For details and the number of required optical elements see Appendix \ref{sec:app_xk}.

\section{Single-photon controlled gates in OAM and path}
\label{sec:controlled}

\begin{figure}
\includegraphics[width=\linewidth]{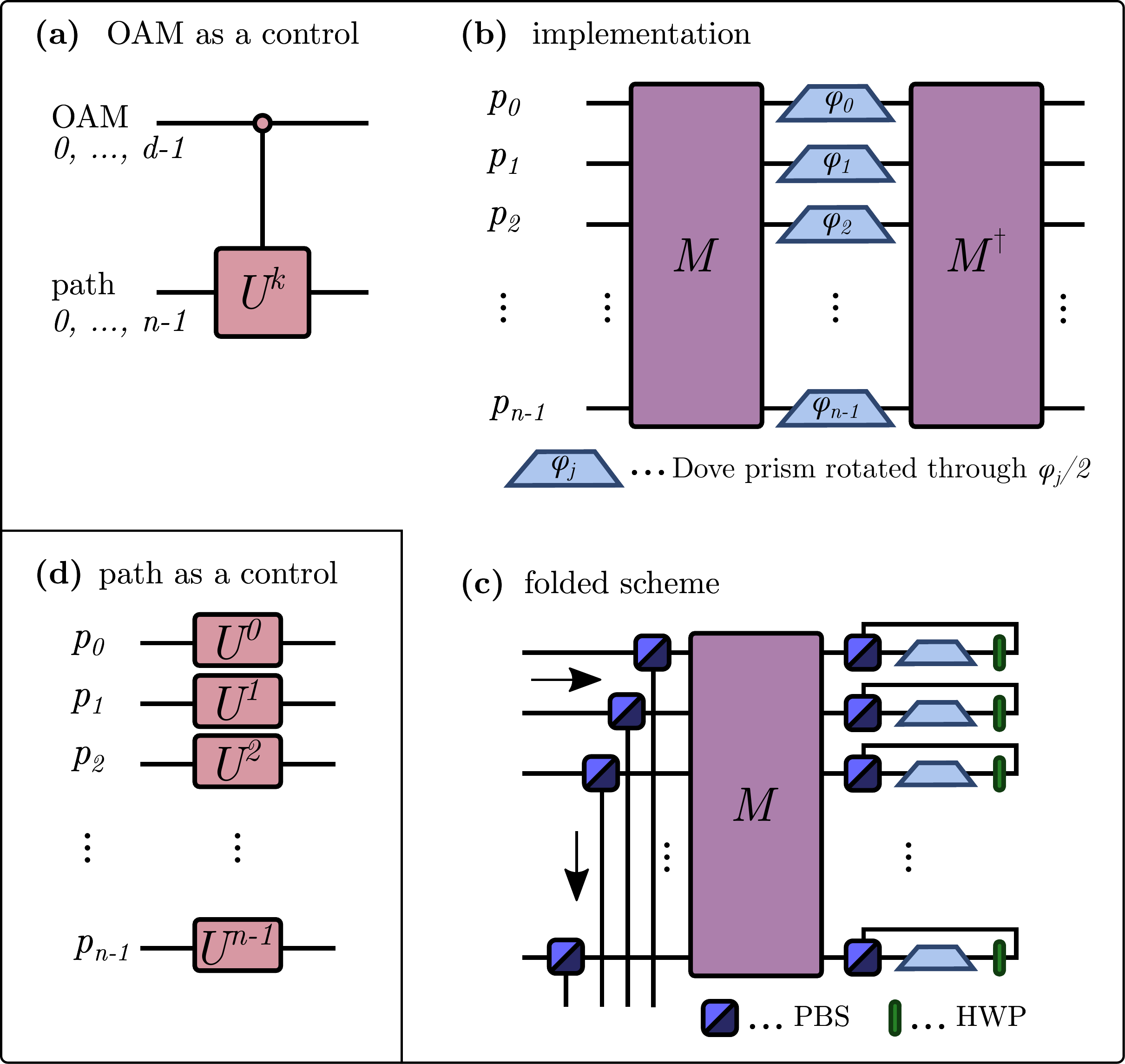}
\caption{Single-photon controlled gates. 
(a) The quantum-circuit representation of a controlled gate $CU$, where the upper line represents a $d$-dimensional OAM qudit and the lower line represents an $n$-dimensional path qudit. (b) The schematics of an actual experimental setup for (a) where individual horizontal lines amount to real propagation paths. The scheme relies on the eigendecomposition of the unitary $U$ as explained in the main text. In each path as many as $d$ different OAM eigenstates can propagate simultaneously. (c) The folded version of (b) reduces the number of optical elements at the cost of using polarization as an additional degree of freedom. A photon enters in the horizontal polarization from the left and leaves the setup in the vertical polarization at the bottom. (d) The schematics of the controlled gate where the control qudit is played by the path modes and individual gates $U^k$ act on OAM.}
\label{fig:oam_control}
\end{figure}

The scheme of preceding section makes use of the path degree of freedom to implement an arbitrary unitary in OAM of a single photon. In this section, we invoke path once more and consider a photon whose state is a superposition of $d$ OAM modes $\ket{k}_O$ propagating along $n$ different paths $\ket{p}_P$. For such states one can study a class of unitaries that consists of controlled gates where the two degrees of freedom play the role of control and target qudits. Single-photon controlled gates acting on spatial modes of light were studied in special cases e.g. in Refs.~\cite{zeng_realization_2016,kagalwala_single-photon_2017}.

\subsection{OAM as a control}

The general action of a controlled operation $CU$ on input states where the OAM and path play the roles of the control and the target qudit, respectively, reads
\begin{equation}
    CU \ket{k}_O\ket{p}_P = \ket{k}_O(U^k \ket{p}_P),
    \label{eq:oam_control}
\end{equation}
where $U$ is a fixed unitary acting on the path degree of freedom. The quantum circuit that corresponds to this operation is depicted in Fig.~\ref{fig:oam_control}(a). An experimental setup that implements this controlled operation can be constructed as follows. First we note that any unitary operation $U$ can be diagonalized, such that
\begin{equation}
    U = M^\dagger \cdot D \cdot M,
    \label{eq:eigdec}
\end{equation}
where $M$ is a unitary matrix and $D$ is a diagonal matrix composed of eigenvalues of $U$. Due to the unitarity, the eigenvalues are of the form $\lambda_j = e^{-i \varphi_j}$ for some real phases $\varphi_j$. From the eigendecomposition formula it directly follows for an arbitrary integer power $k$ that
\begin{equation}
    U^k = M^\dagger \cdot D^k \cdot M.
\end{equation}
The key observation is that $M$ is fixed for all the powers and the diagonal matrix $D^k$ has only complex phases $e^{-i k \varphi_j}$ on its diagonal. These phases can be compared with the action of a Dove prism (DP) rotated through angle $\alpha$, when applied to an eigenstate with $k$ quanta of OAM
\begin{equation}
    \mathrm{DP}_\alpha \ket{k}_O = \exp(-i k 2 \alpha) \ket{-k}_O.
\end{equation}
The extra minus factor in the outgoing OAM eigenstate can be corrected for by an additional mirror.

From the above relations one sees that the controlled unitary $CU$ can be implemented as shown in Fig.~\ref{fig:oam_control}(b). First, an operator $M$ is applied only to the path degree of freedom of the incoming photon; then a series of Dove prisms is utilized, where a Dove prism rotated through $\varphi_j/2$ is placed on the $j$-th path; and finally an inverse operator $M^\dagger$ concludes the operation. That is
\begin{equation}
    CU = M^\dagger \cdot \left( \bigoplus_{j = 0}^{n - 1} \mathrm{DP}_{\varphi_j / 2} \right) \cdot M. \label{eq:cu}
\end{equation}
We thus obtain a passive network of optical elements, where $M$ can be implemented in various ways, such as Reck \emph{et al.} scheme \cite{Reck1994} and its alternatives \cite{Clements2016,Guise2018}. 
Standard Dove prisms have an undesirable impact on the polarization of propagating photons \cite{Moreno04}. Modifications of the Dove prism geometry can nevertheless mitigate this impact considerably \cite{padgett_dove_1999}.

There is one additional feature of the controlled-unitary setup of Fig.~\ref{fig:oam_control}(b) --- suppose that a given unitary $U$ is to act on eigenmodes of the form $\ket{m \, k}$ for $0 \leq k < d$ where $m$ is a fixed integer. The implementation is in this case identical to that in Fig.~\ref{fig:oam_control}(b) except that the Dove prism in the $j$-path is rotated through $\varphi_j/(2 m)$. This general property was noted in Ref.~\cite{isdraila_cyclic_2019} for the special case of the $X$ gate. 

Some savings in resources are possible when a polarization is utilized as an auxiliary degree of freedom. The setup of Eq.~\eqref{eq:cu} has a symmetric structure, where the $M$ operator is applied both before and after the stack of Dove prisms. One can get rid of the second operator to obtain a folded scheme \cite{xgate,isdraila_cyclic_2019}, where $M^\dagger$ is implemented by the backward passage of a photon through the setup for $M$, see Fig.~\ref{fig:oam_control}(c). The forward and backward passage is controlled by the polarization of the photon. Provided that the initial polarization is $H$, the photon traverses both the $M$ module and the Dove prisms as in the original scheme. Then a series of half-wave plates rotates $H$ into $V$ and then all terms in the photon's wavefunction travel backward through $M$. At the end, the terms are reflected out of the setup by an additional series of polarizing beam splitters positioned in front of $M$. Using the Reck \emph{et al.} scheme \cite{Reck1994}, the unfolded scheme can be implemented with $n (n-1)$ beam splitters, $n (n-1)$ phase shifters, and $n$ Dove prisms. The folded scheme requires $n(n+3)/2$ beam splitters (both non-polarizing and polarizing), $n (n-1)/2$ phase shifters, and $n$ Dove prisms. Note that the universal scheme of Eq.~\eqref{eq:universal} in the preceding section can be turned into a folded version in a very analogous way.

\subsection{Examples}

Let us discuss now some special cases of the general scheme \eqref{eq:cu}. The simplest case is when $U$ is itself diagonal. A notable example of such a unitary is the high-dimensional controlled-$Z$ gate, where the $n$-dimensional Pauli $Z_n$ gate is characterized by Eq.~\eqref{eq:zgate}. As follows from Eq.~\eqref{eq:cu}, the $CZ$ gate can be implemented as a mere stack of $n$ properly rotated Dove prisms, where the prism in the $p$-th path is rotated through $\pi p/n$ \cite{oamfft,oamfft2}.

Another special example is the OAM equivalent of the polarizing beam splitter. Such a beam splitter is represented by a high-dimensional controlled Pauli $X$ gate, a high-dimensional generalization of the CNOT gate. The eigendecomposition of the high-dimensional $X$ gate \eqref{eq:xgate} reads $X = F^\dagger \cdot Z \cdot F$, where $Z$ is given in Eq.~\eqref{eq:zgate} and $F$ is the high-dimensional path-only Fourier transform. In the notation of Eq.~\eqref{eq:cu} we have thus $M = F$. As mentioned above, the integer powers of $X$ gate correspond to cyclic permutations with different strides. Thanks to this property the setup of the $CX$ gate can be viewed as a sorter of OAM eigenstates. An experimental implementation of such an OAM sorter based on the aforementioned Fourier relation of $X$ and $Z$ gates was proposed in Ref.~\cite{XuBo2005}. The same idea was then rediscovered 11 years later independently by two groups \cite{PhysRevA.94.033847,Ionicioiu2016}.

Yet another notable example is the Leach interferometer \cite{Leach2002} depicted in Fig.~\ref{fig:xgate}(a), which is usually used as a parity sorter \cite{Leach2004,erhard_experimental_2018}. In this case, operator $M$ corresponds to a single symmetric beam splitter and the diagonal matrix reads $D = \mathrm{diag} (1, \exp(i \alpha))$, where $\alpha$ depends on the intended sorting properties of the Leach interferometer \cite{oamfft}.

\subsection{Path as a control}

For completeness, let us briefly mention the complementary situation to Eq.~\eqref{eq:oam_control} where the path is now the control qudit and OAM is the target qudit
\begin{equation}
    CU \ket{k}_O\ket{p}_P = (U^p \ket{k}_O) \ket{p}_P.
    \label{eq:path_control}
\end{equation}
This case can be formally represented by an abstract quantum circuit akin to that in Fig.~\ref{fig:oam_control}(a), where the roles of control and target qudits are exchanged. On a more practical level, the transformation of Eq.~\eqref{eq:path_control} can be understood as $n$ independent setups, one setup in each path, see Fig.~\ref{fig:oam_control}(d). 
The easy approach how to implement such a controlled operation is to use a swap operator, discussed in detail in the following section, that exchanges the role of the path and OAM and sandwich the setup of Fig.~\ref{fig:oam_control}(b) between two such swaps. The advantage of such a scheme, when compared to the naive scheme of Fig.~\ref{fig:oam_control}(d), is the reduction in the amount of resources. In the general case, one requires $O(d^2 n)$ elements to implement the naive scheme, whereas roughly $O(d^2 + n \log(n))$ elements are required by the scheme based on swaps and Fig.~\ref{fig:oam_control}(b). Analogously to the previous section, also in the present scenario we can construct the folded scheme. The polarizing beam splitters and half-wave plates reroute the terms of the photon wave function such that the initial polarization $H$ is midway through the setup changed into $V$ and the setup is propagated backward.

The circuit of Fig.~\ref{fig:oam_control}(d) can also be seen from the multiple-photon point of view where each path is occupied by exactly one photon. Even in such a case the setup can be implemented by sandwiching the scheme of Fig.~\ref{fig:oam_control}(b) between two swaps, where individual photons share some propagation paths. This point of view leads us to study the parallelization of the scheme in the following section, where the same local unitary is applied to multiple photons.

\section{Parallelization}
\label{sec:parpaths}

\begin{figure*}
\includegraphics[width=0.95\linewidth]{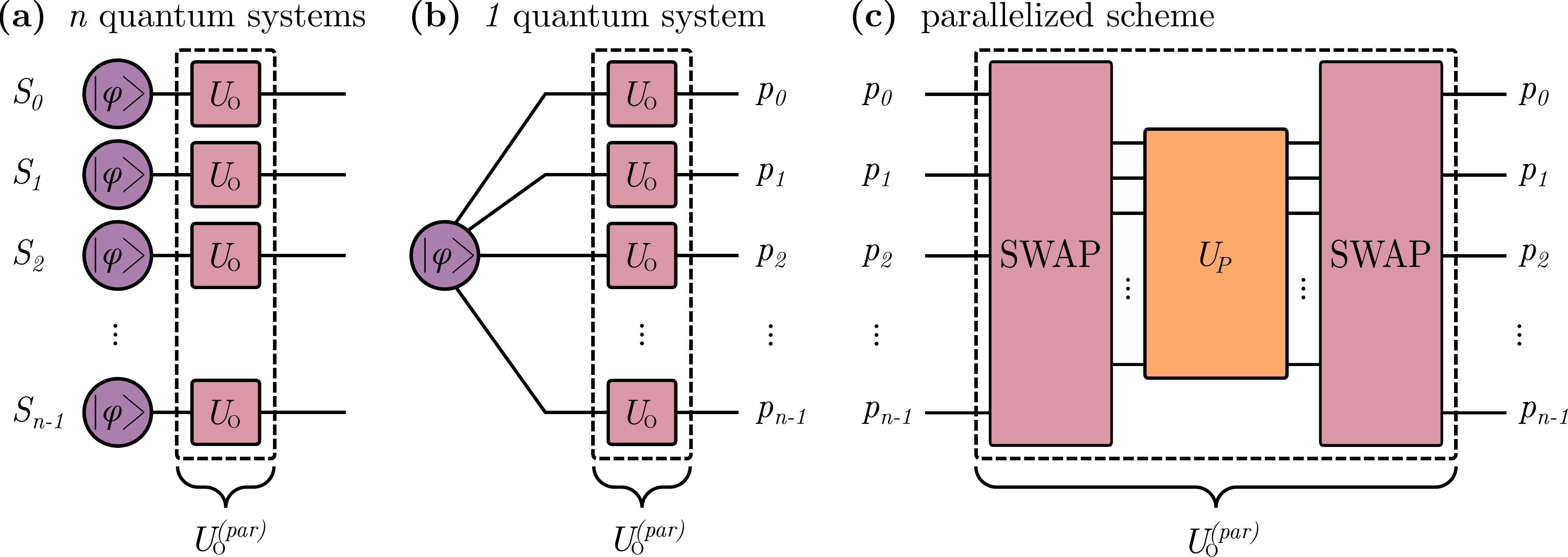}
\caption{Parallelization schemes. Some computational tasks require the application of the same unitary operation to multiple path modes. (a) One can consider $n$ different quantum systems $S_j$, each in a separate path and subjected to a local unitary $U_O$. (b) A qualitatively different scenario is that of a single quantum system that propagates along a superposition of multiple paths $p_i$ and is subjected to an operation $U$ that acts only on its internal degree of freedom (such as OAM). (c) The series of identical devices $U_O$ acting on the internal degree of freedom can be replaced by a parallelized scheme of Eq.~\eqref{eq:paral}.}
\label{fig:naive}
\end{figure*}

\subsection{General case}

In the domain of classical computation, a truly large-scale deployment is allowed by various parallelization techniques and the same can also be expected in the quantum domain. In general, a large complex computational task is split into smaller parts, each of which is computed by a separate computational core. The simplest case is when the task consists of multiple identical subtasks. One arrives at the scenario of Fig.~\ref{fig:naive}(a), where each subtask is represented by a local unitary $U_O$ acting on one photon. The resulting collection of unitaries $U_O$ can be viewed as a single parallelized operation $U_O^{\mathrm{(par)}}$ acting on many photons. Such a stack of identical gates is henceforth referred to as the \emph{naive approach}.

It is noteworthy to point out that the same setup of multiple local unitaries emerges in a qualitatively different scenario when a \emph{single} photon propagates in a superposition of multiple paths and one wants to apply operation $U$ only to its internal degree of freedom, such as OAM. In the experimental realization, $U$ is implemented as a series of identical setups with one setup $U_O$ in each path, see Fig.~\ref{fig:naive}(b). This is reminiscent of the scheme in Fig.~\ref{fig:oam_control}(d) except that the unitary in each path is the same.

In the naive approach, $U_O^{\mathrm{(par)}}$ requires a number of elements that scales linearly with the number of systems. The same task of simultaneous application of $U$ on $n$ paths can be nevertheless achieved with just a single device, as shown in Fig.~\ref{fig:naive}(c). The key role in this approach is played by the swap operator, whose action on input states reads
\begin{equation}
    \swap(\ket{m}_O\ket{p}_P) = \ket{p}_O\ket{m}_P,
    \label{eq:swap}
\end{equation}
where $\ket{m}_O$ denotes internal mode $m$ and $\ket{p}_P$ stands for the $p$-th propagation mode. In the scheme, a swap first exchanges the roles of internal and path modes. The path-encoded implementation $U_P$ of the desired unitary $U$ is then applied to the photon(s). This operation transforms the path modes according to $U$ and leaves the internal modes unaffected. Even though this property may not be satisfied in general, in many cases this is indeed the case as $U_P$ can be constructed only with beam splitters and phase shifters \cite{Reck1994}, which leave e.g. OAM, polarization, or frequency of photons unaffected\footnote{In the universal scheme in Sec.~\ref{sec:universal} this was not an issue as all the paths contained only the fundamental internal OAM mode.}. In the third stage, a swap is applied again in order to give the internal and path modes their original meaning. As a result, the internal modes in each path are transformed according to $U$. This procedure is summarized by the formula
\begin{equation}
    U_O^{\mathrm{(par)}} = \swap^{-1} \cdot U_P \cdot \swap, \label{eq:paral}
\end{equation}
which we henceforth refer to as the \emph{parallelized scheme}. The parallelized scheme is clearly a generalization of the approach used to construct an arbitrary unitary in OAM in section \ref{sec:universal}. The sorter from Eq.~\eqref{eq:sorter} can be understood as a special case of the swap operator in Eq.~\eqref{eq:swap} for $p = 0$.

Even though the relation \eqref{eq:paral} holds for any internal degree of freedom, it is not obvious how to implement efficiently the $\swap$ operator in a general case. For the case of OAM and path we can utilize the efficient design of Ref.~\cite{oamfft} whose detailed structure is presented in section \ref{sec:paralxk}. The path-encoded unitary $U_P$ can be implemented using Reck \emph{et al.} scheme \cite{Reck1994} and the parallelized setup of Eq.~\eqref{eq:paral} is thus made of only conventional optical elements. Moreover, each beam splitter in the Reck \emph{et al.} scheme has to be supplemented by two extra mirrors, such that the sign of OAM eigenstates is unaffected by the reflection off the beam splitter's interface \cite{oamfft}.

\subsection{Scaling of resources}
\label{sec:scaling}

The simultaneous $n$-fold application of unitary $U_O$ can be in the naive approach implemented with $n$ separate setups. To quantify the improvement brought by the use of the parallelized scheme of Eq.~\eqref{eq:paral}, we introduce the ratio of the number of beam splitters required by the parallelized and naive schemes
\begin{equation}
    r(n, d) \coloneqq \frac{\suptext{N_O}{par}(n,d)}{n \, N_O(d)}.
    \label{eq:r}
\end{equation}
The smaller this ratio, the better the parallelized scheme is when compared to the naive implementation. One can consider similar ratios also for other optical elements, with similar results to those presented below. Some general statements about the efficiency of the improved scheme can be derived even for an unspecified unitary operation. According to Reck \emph{et al.} scheme at most $N_P(d) = d(d-1)/2$ beam splitters are necessary to implement an arbitrary path-encoded unitary $U$. For large enough dimensions the ratio $r_{\mathrm{Reck}}$ scales as
\begin{equation}
    r_{\mathrm{Reck}}(n, d \gg n) \sim \frac{1}{n},
    \label{eq:r1}
\end{equation}
which can be interpreted such that the parallelized scheme uses effectively only one setup for $U_P$ instead of $n$ identical setups for $U_O$. The parallelized scheme thus scales in this general case linearly better than the naive approach. For more details refer to Appendix \ref{sec:swap}.

For specific classes of unitaries one obtains different scaling estimates. The extreme case is represented by unitaries that correspond to mere permutations of modes. For those, we get $N_P(d) = 0$ as no beam splitters are necessary to permute paths. The ratio for both $n > d$ and $n \leq d$ cases then scales as
\begin{equation}
    r_{\mathrm{perm.}}(n, d) \sim \frac{\log_2(d)}{2 n} + \frac{\log_2(n)}{4d}.
    \label{eq:rPerm}
\end{equation}
Except for extreme cases the parallelized scheme is again more efficient than the stack of $n$ independent setups. The aforementioned complexity estimates are based on the assumption that no simplification of the resulting setup is possible. Nevertheless, the scaling may differ when the implementation of the parallelized scheme can be simplified by removing superfluous elements. We exemplify this reduction in section \ref{sec:paralxk} for integer powers of the high-dimensional $\xgate$~gate, which form a special class of permutations. Another situation when the above general estimates do not hold is when other than the brute-force implementation from Eq.~\eqref{eq:universal} is used to construct $U_O$. For instance, the $d$-dimensional Fourier transform of the OAM eigenstates of a single photon can be implemented using only $N_O(d) \sim \sqrt{d} \log(d)$ beam splitters \cite{oamfft2}. Even though the Fourier transform is a much more complex operation than a mere permutation, the scaling of the corresponding number of beam splitters is basically identical to \eqref{eq:rPerm}.

Additional savings in resources are possible when a polarization is utilized in the scheme of Eq.~\eqref{eq:paral} to arrive at the folded setup in a way completely analogous to that described in section \ref{sec:controlled}. The number of beam splitters is then reduced approximately by the number of beam splitters required to construct the removed swap operator.

The parallelized scheme of Eq.~\eqref{eq:paral} reduces the total number of elements, but the number of elements that each photon has to traverse on average increases. Another relevant issue when assessing the performance of the setup are thus losses accompanying the transformation $U_O^{\mathrm{(par)}}$. Even though the detailed analysis of the role of losses lies beyond the scope of the present paper, in Appendix \ref{sec:losses} a simplified discussion is presented. It turns out that the overall transmittance per photon of the parallelized scheme is decreased by a factor of $T^{2 \log_2(d)}$ when compared to the naive scheme, where $T$ quantifies the effective mean transmittance of each optical element in the setup. The losses for both the naive and parallelized schemes are otherwise comparable to schemes that implement purely path-encoded unitaries, such as the scheme of Ref.~\cite{Clements2016}.

\subsection{Parallelized Pauli gates}
\label{sec:paralxk}

\begin{figure*}
\includegraphics[width=0.95\linewidth]{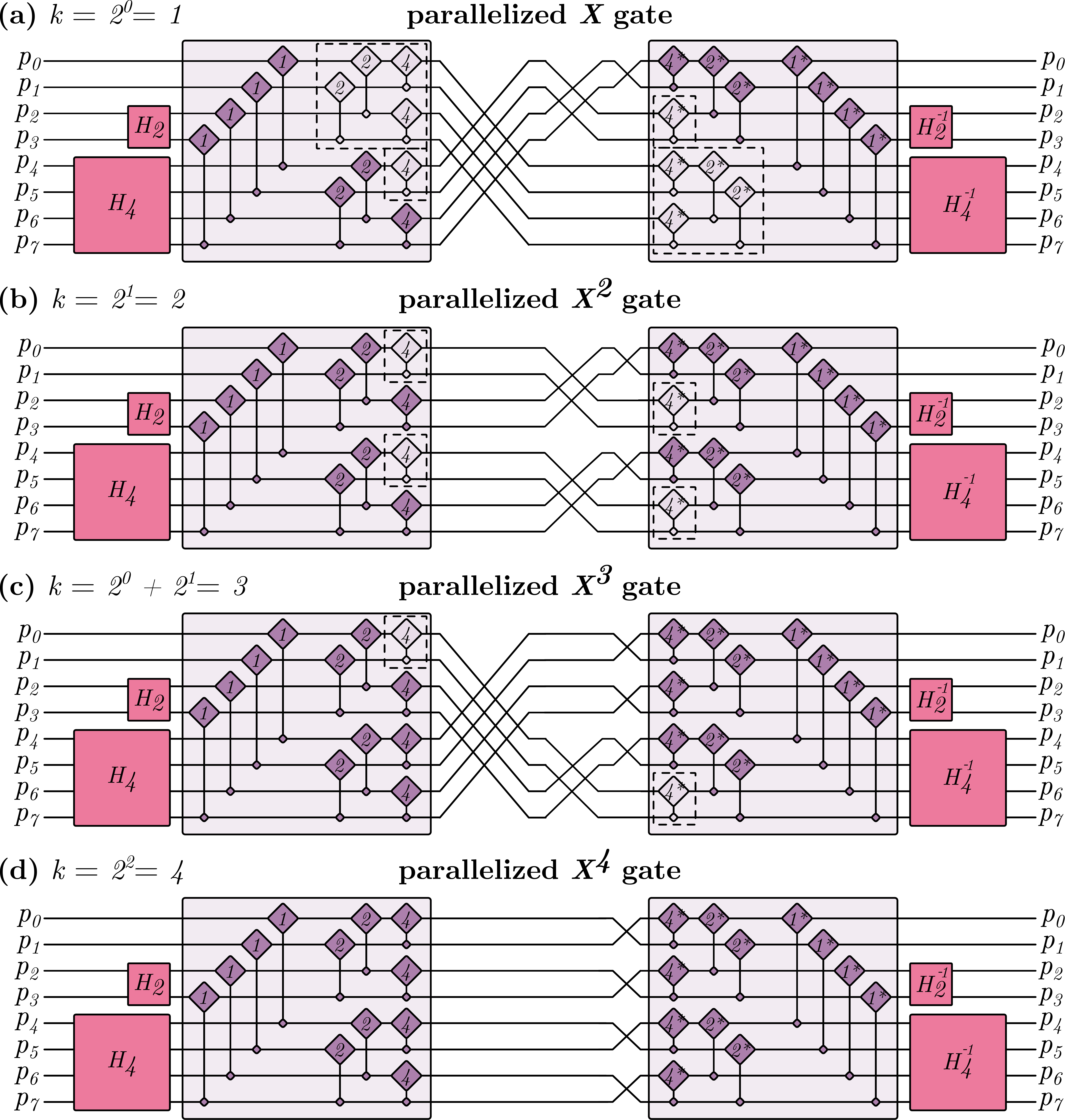}
\caption{Explicit forms of the parallelized scheme for $\xk$ gates. The parallelized scheme, exemplified for $d = n = 8$ and $1 \le k \le d/2$, consists of a path permutation sandwiched between two swaps. The swap operator comprises a network of OAM exchangers, whose structure is depicted in Fig.~\ref{fig:xgate}, and a series of $H$ blocks (in the present case there are two such blocks per one swap). The presented schemes can be simplified by removing the exchangers that do not affect the final state. These are drawn in faded colors and enclosed in dashed boxes. The input paths of the swap operator should be permuted to comply with formula \eqref{eq:swap}. As this path permutation does not affect our discussion, we omit it in the figure for clarity. (a) Parallelized $\xgate$ gate. (b) Parallelized $\xgate^2$ gate. (c) Parallelized $\xgate^3$ gate. (d) Parallelized $\xgate^4$ gate.}
\label{fig:xkgate_paral}
\end{figure*}

\begin{figure}
\includegraphics[width=\linewidth]{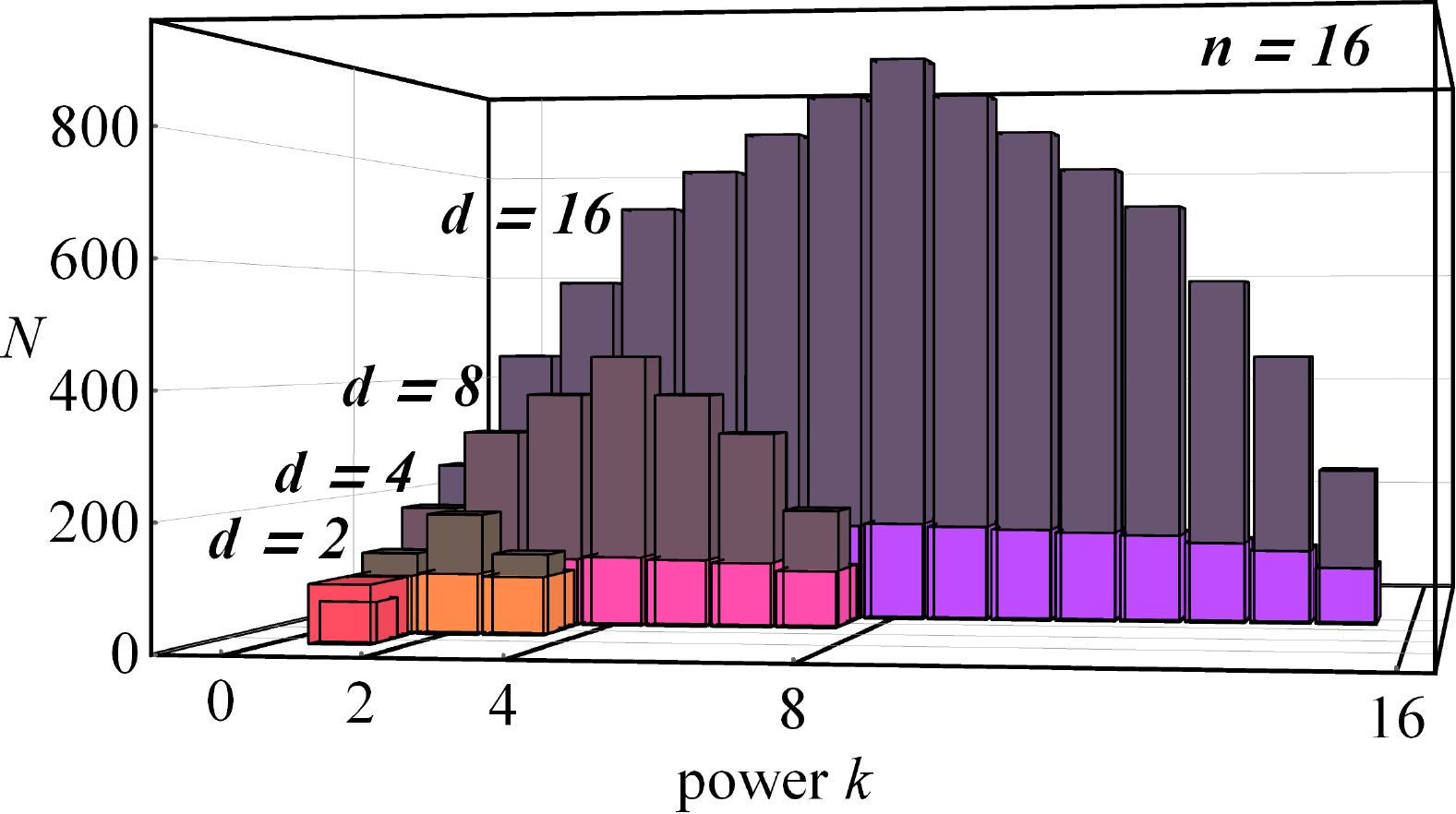}
\caption{The number of beam splitters $N$ in the naive and parallelized implementations of $d$-dimensional $\xk$~gates. The gates act on $n=16$ propagations paths in dimensions $d = 2, 4, 8, 16$ and $1 \leq k \leq d - 1$. The naive implementation, represented by dark bars, requires as many as 960 beam splitters for $d = 16$ and $k = 8$. On the contrary, the parallelized implementation, represented by bright bars, needs only 162 beam splitters in such a case.}
\label{fig:numBS}
\end{figure}

In this section, we discuss the parallelization of Pauli gates and compare their scaling properties with the naive approach. As mentioned in section \ref{sec:pauli}, the implementation of the $d$-dimensional Pauli $Z$ gate is especially simple---a single rotated Dove prism will do. The $Z$ gate in OAM is thus an example of a gate where the parallelization does not actually bring any advantage. This is no longer true though for the $X^k$ gates.

One can parallelize the $\xgate$~gate by starting from the setup in Eq.~\eqref{eq:paral}, where $U_P$ is the path-encoded $\xgate$~gate, and then removing all the redundant optical elements. To see which elements are not necessary, let us take a close look at the internal structure of the swap operator demonstrated in Fig.~\ref{fig:xkgate_paral}(a). The swap consists of two functionally different parts \cite{oamfft}---one $E$ block and a series of $H$ blocks of increasing size. The $E$ block is a network of exchangers $\exch{k}$ of increasing orders of the form $k = 2^l$ and is shown explicitly for each swap in Fig.~\ref{fig:xkgate_paral}. The structure of $H$ blocks is not of interest in our discussion and can be found in Ref.~\cite{oamfft}. The removal of redundant exchangers in the case of the parallelized $\xgate$~gate is depicted in Fig.~\ref{fig:xkgate_paral}(a) explicitly for the special example of $n = d = 8$. Analogously to the procedure of section \ref{sec:pauli}, there are many instances where an OAM exchanger $\exch{k}$ is followed by its inverse $\exch{k}^{-1}$. These exchangers can be removed without affecting the final state. In a completely analogous way one also proceeds for the parallelized integer powers, the $\xk$~gates. One again starts from the scheme in Eq.~\eqref{eq:paral}. This step for $n = d = 8$ is shown in Figs.~\ref{fig:xkgate_paral}(b), (c), and (d) for all $\xk$~gates with $k \le d/2$.

The number of beam splitters in the parallelized scheme is shown in Fig.~\ref{fig:numBS} for $n=16$ propagation paths and dimensions $d \leq n$. For comparison, the naive approach that involves $n$ copies of the non-parallelized $\xk$~gate is also shown. To put these numbers into context, we note that in Ref. \cite{Zhongeabe8770} an experiment has been reported recently where 50 polarizing beam splitters and a bulk interferometer representing 300 beam splitters were used. Even though the naive approach exceeds these numbers already for $d=8$, the parallelized scheme allows for the construction of an arbitrary power of the $\xgate$~gate even for $d=16$. At most 162 beam splitters are required in such a case. Although this number is still rather formidable from the present technology point of view, the savings provided by the parallelized schemes are clearly visible. In Appendix \ref{sec:paral_xk} a more detailed discussion of the number of elements is presented.

\section{Periodicity}
\label{sec:periodicity}

Each implementation scheme discussed so far comes with one neat feature --- they are periodic in OAM. When working with OAM degree of freedom, one has to define the subspace of eigenstates in which the operations are to be carried out. When one sets the dimension to $d$, one possible choice of the OAM subspace consists of eigenstates $\{\ket{0}, \ldots, \ket{d-1}\}$. Let us denote the subspace spanned by these eigenstates with $\hilb_0$. Another choice of eigenstates can be $\{\ket{d}, \ldots, \ket{2d-1}\}$ or $\{\ket{a \, d}, \ldots, \ket{(a+1)d-1}\}$ for a general $a \in \mathbb{Z}$. Let us denote the subspace spanned by the latter eigenstates by $\hilb_a$. Consider a unitary operation $U$ defined on subspace $\hilb_0$ by formula
\begin{equation}
    U \ket{i} = \sum_{j=0}^{d-1} U_{j,i} \ket{j},
\end{equation}
where $U = (U_{i,j})$ and $0 \leq i, j < d$. For dimensions of the form $d = 2^M$, the naive implementation $U_O$ \eqref{eq:universal} of unitary $U$ acts identically on each subspace $\hilb_a$ for \emph{any} $a \in \mathbb{Z}$, not only on the fundamental subspace $\hilb_0$, such that
\begin{equation}
    U \ket{i + a \, d} = \sum_{j=0}^{d-1} U_{j,i} \ket{j + a \, d}.
    \label{eq:oam_paral}
\end{equation}
This property was noted for the case of the high-dimensional $\xgate$~gate in Ref.~\cite{xgate}, but any power of the $\xgate$~gate constructed in section \ref{sec:pauli} has this property also\footnote{The parallelized scheme exhibits the same periodicity property with one modification. As follows from the construction of the swap operator, when $n > d$ the values of input OAM eigenstates have to be chosen like $\ket{i \, n/d}$. The formula \eqref{eq:oam_paral} is then modified such that $i \to i n /d$ and $j \to j n/d$ where we assume that not only the OAM-space dimension but also the number of paths is a power of two.}. The periodicity in OAM is a result of the modulo property of the OAM sorter and the swap operator \cite{oamfft,Palici_2020}. For details refer to Appendix \ref{sec:oamper}.

An interesting issue is that of the periodicity of the controlled-unitary setup of Eq.~\eqref{eq:cu}. When the eigenvalues of $U$ are of the form $\lambda_k = \exp(2 \pi i (a_k/b_k))$ for some $a_k, b_k \in \mathbb{Z}$, the whole scheme is periodic with the period that does not exceed the product $b_0 b_1 b_2 \ldots b_{n - 1}$. Unitaries whose eigenvalues have phases that are irrational multiples of $2 \pi$ do not display any periodic behaviour. In practice though, any real number can be approximated by a rational number and the periodicity is effectively restored.

The periodicity of the presented setups may be seen as another parallelization feature --- the parallelization in OAM. The OAM degree of freedom allows for the generation of very-high-dimensional states. One can thus consider a large OAM Hilbert space $\hilb$ composed of $n$ subspaces $\hilb_a$, each of dimension $d$, such that $\hilb$ is a direct sum of the form
\begin{equation}
    \hilb = \hilb_1 \oplus \hilb_2 \oplus \ldots \oplus \hilb_n.
\end{equation}
A single photon with $(nd)$-dimensional state $\ket{\psi} \in \hilb$ can therefore be seen as representing a sum of $n$ different $d$-dimensional qudits $\ket{\psi_a} \in \hilb_a$. When the manipulation of this state is done via operations discussed in preceding sections, each qudit $\ket{\psi_a}$ is due to Eq.~\eqref{eq:oam_paral} manipulated independently of all the others. One photon can thus in effect carry $n$ different qudits in parallel. This framework can be seen as \emph{subspace multiplexing}, where several OAM eigenmodes together carry a given quantum state in each subspace. Subspace multiplexing could be used in the free space communication or in the quantum computation with single photons.

Even though in theory there is no limit on the largest possible number of OAM quanta and, therefore, the number of subspaces the operation $U_O$ can act on, from the physical perspective there is a limit. The spatial extent of the photon's wavefunction increases with the number of OAM quanta and so for large OAM values the eigenstate becomes macroscopic and unwieldy for manipulation \cite{Campbell_12,Shen_2013,Fickler13642}.

\section{Conclusion}
\label{sec:conclusion}

We study the manipulation of orbital angular momentum of light with the help of interferometric networks. The networks consist of conventional optical elements and spiral phase plates, which add or subtract an integer number of OAM quanta. Importantly, no use of holograms with complex phase profiles is made, as is the case for instance in the multi-plane light conversion method \cite{Morizur10,Labroille2014,Fontaine2019,Brandt20}. The interferometric scheme for implementing an arbitrary unitary operation on single photons is presented, which can be viewed as the OAM counterpart of Reck \emph{et al.} scheme. To exemplify the argument, we construct explicitly the setups for $\xk$~gates for dimensions of the form $d = 2^M$ and derive precise analytical formulas for the number of employed optical elements.

Another interferometric scheme introduced is that for implementation of single-photon controlled gates, where the OAM and path play the role of either the control or target qudits. It turns out that several reported results of other authors are special cases of the scheme.

The last class of interferometric networks under consideration is the one that allows for simultaneous application of the same unitary on multiple OAM states. This parallelized scheme can find applications e.g. in multiplexed communication channels, where the internal modes of photons are used as carriers of information. The savings in resources for a general unitary offered by the parallelized scheme scale approximately linearly with the number $n$ of involved parties, i.e., the naive approach requires roughly $n$ times more elements than the parallelized approach. The parallelized versions of Pauli gates are constructed as concrete examples.

The networks presented in the text display periodicity in OAM, which can be used to implement subspace multiplexing where one device acts on many disjoint OAM subspaces of the same photon in the same way. When multiple independent pieces of information are encoded in the OAM of a single photon, our scheme applies the same unitary to each piece separately. The periodicity follows directly from the sorting properties of Leach interferometers in dimensions that are powers of two. This feature is unique to the interferometric implementation we use in this paper and cannot be imitated by alternative techniques that implement an OAM unitary using holograms with complex phase profiles \cite{Berkhout2010,Labroille2014,Brandt20}.

The above results are in principle applicable not only to single photons but also to classical beams of light carrying orbital angular momentum. It would be interesting to see whether more properties of the networks similar to those studied in the text can be found.

We would like to thank Robert Kindler and Johannes Pseiner for their valuable comments on the early version of the manuscript. This work was supported by the Austrian Academy of Sciences and the University of Vienna via the QUESS project (Quantum Experiments on Space Scale).

\appendix

\section{Number of elements}
\label{sec:num_of_elems}

\subsection{General unitary}

The path-encoded unitary $U_P$ as well as the OAM sorter are implemented as networks of many interferometers. Beam splitters thus play an important role in their construction. In the following, we will estimate the complexity of the OAM implementation $U_O$ by counting the beam splitters used in its construction. A similar discussion can also be done for other optical elements. Let $N_O(d)$ be the number of beam splitters required in the scheme of Eq.~\eqref{eq:universal}. If $N_P(d)$ is the number of beam splitters that are necessary to implement $U_P$, then
\begin{eqnarray}
    N_O(d) = N_P(d) + 2 N_{\sorter{}}(d),
    \label{eq:numbs}
\end{eqnarray}
where
\begin{equation}
     N_{\sorter{}}(d) = 2 \, (d-1)
     \label{eq:sorter_numbs}
\end{equation}
is the number of beam splitters that implement the OAM sorter in dimension $d$ \cite{oamfft}. The upper bound for $N_P(d)$ is provided by Reck \emph{et al.} scheme, for which $N_P(d) = d(d-1)/2$.

\subsection{$\xk$ gates}
\label{sec:app_xk}

The number of beam splitters required in a scheme for the $d$-dimensional $\xgate$~gate, where $d = 2^M$, is equal to \cite{xgate}
\begin{equation}
    N_{\xgate}(d) = 4 \log_2(d).
    \label{eq:xgate_simple_scaling}
\end{equation}
Following the analysis in the main text, it is not hard to see that the number of beam splitters necessary to implement $\xk$, where $d = 2^M$ and $k = 2^m$, is
\begin{eqnarray}
    N_{\xgate}(d, k = 2^m) & = & k \, N_{\xgate}(d/k) + 2 \, N_{\sorter{}}(k) \\
    & = & 4 \, \left(k \, \log_2 \left(\frac{d}{k} \right) + k - 1 \right).
    \label{eq:xk_simple}
\end{eqnarray}
It is easy to observe that for $k = d/2 = 2^{M-1}$ there are no exchangers to be removed from the naive scheme, see Fig.~\ref{fig:xgate}(g), since the path permutation affects all the outermost exchangers in both OAM sorters. Indeed, in such a case $N_{\xgate}(d, k = d/2) = 2 \, N_{\sorter{}}(d)$. On the other hand, for $k = 1 = 2^0$ the number of required beam splitters is minimal and \eqref{eq:xk_simple} coincides with $N_{\xgate}(d)$ \eqref{eq:xgate_simple_scaling}.

The structure of the network for powers $k$ that are not powers of two is more complicated. Even though the same procedure as delineated in the main text can be followed, we refrain from describing it in detail and present the resulting number of beam splitters retained in the simplified setup. It turns out that this number for $\xk$~gates in dimension $d = 2^M$ and for general powers $1 \le k \le d/2$ is equal to
\begin{equation}
    N_{\xgate}(d,k) = 4 \, \left(k \, \log_2 \left(\frac{d}{2^{m+1}} \right) + 2^{m+1} - 1\right),
    \label{eq:xk_nonpower_k}
\end{equation}
where $m$ is an integer such that $2^m \le k < 2^{m+1}$. For $k = 2^m$ we recover formula \eqref{eq:xk_simple}.

From the structure of the OAM exchanger, Fig.~\ref{fig:xgate}(a), and the fact that the resulting setup for any $\xk$ gate consists only of the OAM exchangers, it is clear that the exact same formula \eqref{eq:xk_nonpower_k} applies also to the number of employed Dove prisms and holograms. For the number of mirrors we get twice as large a number and there is no need for phase shifters.

\subsection{Parallelized scheme}
\label{sec:swap}

\begin{figure}
\includegraphics[width=\linewidth]{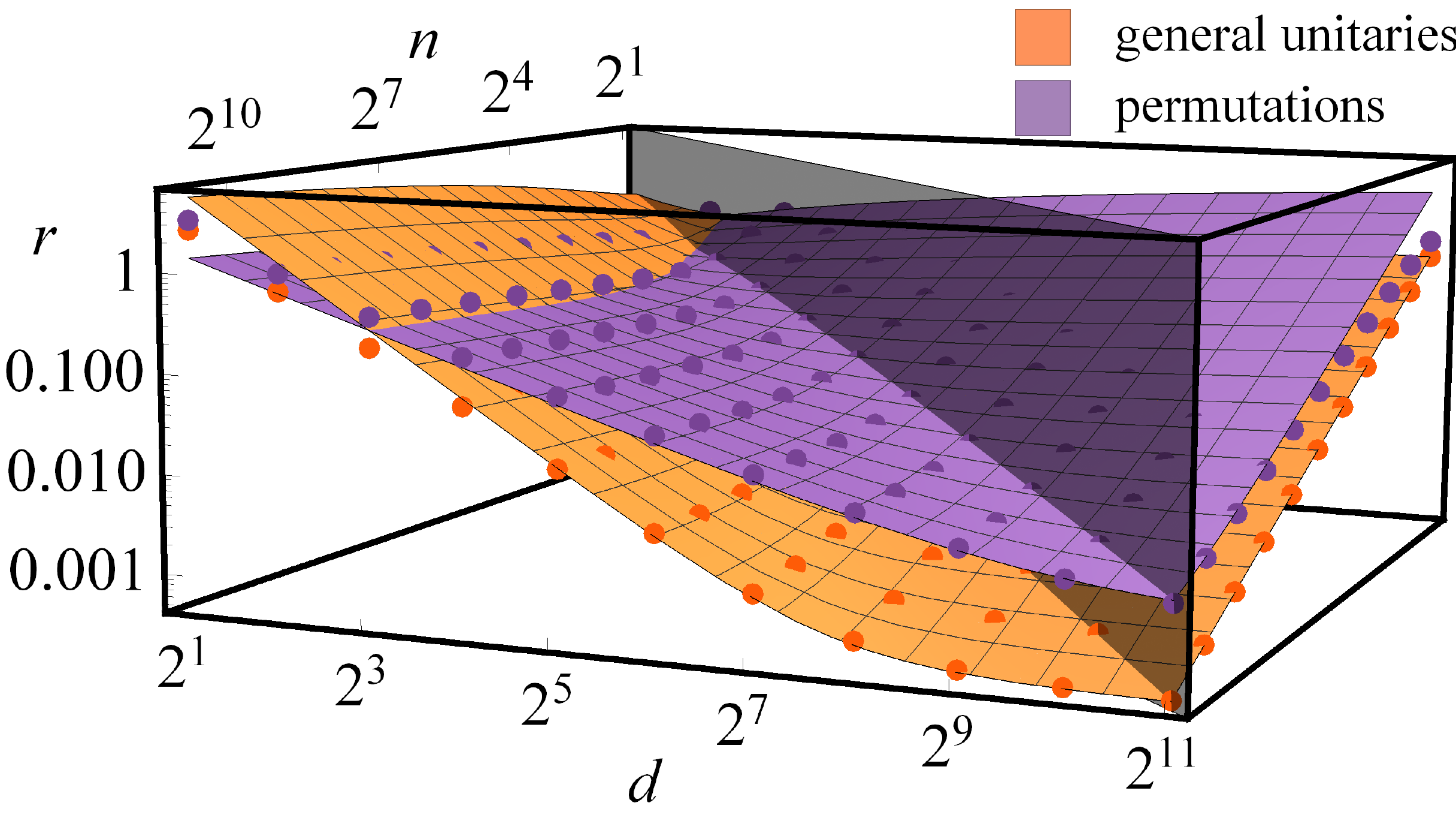}
\caption{The ratio \eqref{eq:r} plotted as a function of the number of paths $n$ and the OAM dimension $d$. Orange dots represent exact values of $r_{\mathrm{Reck}}$ for general unitaries implemented with Reck \emph{et al.} scheme \cite{Reck1994}. Violet dots correspond to the ratio $r_{\mathrm{perm.}}$ for permutations, which form a special subset of unitaries. The orange surface is given by formula \eqref{eq:rReck} that expresses the approximate scaling of ratio $r_{\mathrm{Reck}}$. The violet surface is likewise given by formula \eqref{eq:rPerm} derived for permutations. The bottom right corner of the plot corresponds to $n = d = 2048$, for which $r_{\mathrm{Reck}} \sim 10^{-4}$ for general unitaries and $r_{\mathrm{perm.}} \sim 10^{-3}$ for permutations. The gray plane divides the cases with $n > d$ and $n \leq d$.}
\label{fig:plot_ratio}
\end{figure}

Let us denote by $\suptext{N_O}{par}$ the number of beam splitters employed in the parallelized scheme of Eq.~\eqref{eq:paral}. Analogously to Eq.~\eqref{eq:numbs} we obtain
\begin{eqnarray}
    \suptext{N_O}{par}(n,d) = N_P(d) + 2 \, N_{\swap}(n,d),
\end{eqnarray}
where $N_{\swap}(n,d)$ is the number of beam splitters used to implement the swap operator with $n$ input and $d$ output paths. Due to the structure of the swap operator \cite{oamfft} we have to discuss the case with $n \leq d$ and that with $n > d$ separately. When both $n$ and $d$ are powers of two and $n \leq d$, the number of beam splitters that implement the swap operator is given by \cite{oamfft2}
\begin{equation}
        N_{\swap}(n,d) = \frac{n}{2} \log_2(n) + d \log_2(n) - 3 n + 2 d + 1.
    \label{eq:numswap_nsd}
\end{equation}
In the opposite case with $n \geq d$ one obtains
\begin{equation}
    N_{\swap}(n,d) = \frac{n}{2} \log_2(n) + d \log_2(d) + n - 2 d + 1.
    \label{eq:numswap}
\end{equation}
When $n$ or $d$ are not powers of two, we construct the swap with $2^r$ input and $2^s$ output paths, where $r$ and $s$ are such that $2^{r-1} < n \leq 2^r$ and $2^{s-1} < d \leq 2^s$. Formulas \eqref{eq:numswap_nsd} and \eqref{eq:numswap} then represent upper bounds on the number of utilized beam splitters.

To compare the performance of the parallelized scheme with the naive approach, let us use the ratio $r$ defined in Eq.~\eqref{eq:r} and consider a general unitary implemented with Reck \emph{et al.} scheme \cite{Reck1994}. For this scheme the ratio \eqref{eq:r} scales roughly as
\begin{equation}
    r_{\mathrm{Reck}}(n, d) \sim \frac{1}{n} + \frac{2 \log_2(n)}{d^2}.
    \label{eq:rReck}
\end{equation}
Even though the number of beam splitters in the swaps differs for the case of $n \leq d$ and that of $n > d$, the scaling \eqref{eq:rReck} holds approximately for both of them. A sample of exact values of ratio $r_{\mathrm{Reck}}$ as well as the asymptotic behavior are depicted in Fig.~\ref{fig:plot_ratio}. Permutations serve as a counterpart of this general scheme as far as the number of beam splitters is considered, see $r_{\mathrm{perm.}}$ in Eq.~\eqref{eq:rPerm}. For comparison, a sample of exact values of $r_{\mathrm{perm.}}$ as well as the asymptotic behavior \eqref{eq:rPerm} are also depicted in Fig.~\ref{fig:plot_ratio}.

As for the other optical elements involved in the parallelized setup built using Reck \emph{et al.} scheme, the following estimates apply. The path-only unitary implementation requires $O(d^2)$ beam splitters and $O(d^2)$ phase shifters. To build the swap operators there are $O(n \log_2(n))$ beam splitters, $O(n)$ phase shifters, $O(n \log_2(n))$ holograms, and $O(n)$ Dove prisms necessary, provided that $n > d$ \cite{oamfft2}. When $n \leq d$, the estimates only depend all on $d$, not on $n$.

\subsection{Parallelized $\xk$ gates}
\label{sec:paral_xk}

The resulting number of beam splitters required to implement the parallelized version of the $\xgate$~gate in dimension $d = 2^M$ for $n = 2^K$ paths is equal to
\begin{equation}
    N_{\xgate}^{(\mathrm{par})}(n, d) = n \log_2(n) + 2 \, n - 2,
    \label{eq:xgate_scaling}
\end{equation}
provided that $n \geq d$. This formula does not depend on the dimension $d$, only on the number of paths. The naive approach consisting in stacking $n$ non-parallelized schemes \eqref{eq:xgate_simple_scaling} would require $N_{\xgate}(d) \, n = 4 n \log_2(d)$ beam splitters. The saving in resources is thus approximately equal to
\begin{equation}
    r_{\xgate}(n, d) \approx \frac{\log_2(n)}{4 \log_2(d)}
    \label{eq:rparx}
\end{equation}
for large enough dimensions $d$ and number of paths $n \geq d$. When the number of paths is approximately equal to the dimension, the ratio above approaches a constant factor of $1/4$ and the parallelization of Eq.~\eqref{eq:paral} provides a moderate improvement over the naive approach. When $n \leq d$, the formula \eqref{eq:xgate_scaling} is modified, but even then the improvement resulting from the parallelized scheme is rather moderate.

By calculations analogous to those for the non-parallelized powers of $\xk$ gates, the number of beam splitters retained in the final implementation of the parallelized schemes for an arbitrary $k \leq d/2$ turns out to be
\begin{equation}
    N_{\xgate}^{(\mathrm{par})}(n, d, k) = n \log_2(n) + 2 n - 4 k + 2 + 2 d \left( \frac{k}{2^m} + m - 1\right),
    \label{eq:xkgate_scaling}
\end{equation}
where $m$ in an integer such that $2^m \le k < 2^{m+1}$ and where we assume $n \geq d$. This expression simplifies for $k = 1$ into the formula \eqref{eq:xgate_scaling} derived for the parallelized version of the $\xgate$~gate. A similar discussion can also be done for other optical elements with similar results and for $n < d$.

When we compare the scaling for the naive approach utilizing $n$ identical copies of the $\xk$~gate and the parallelization of Eq.~\eqref{eq:paral}, we obtain a scaling ratio that approaches
\begin{equation}
    r_{\xgate}(n, d, k) \sim \frac{1}{4 \, k} \frac{\log_2(n)}{\log_2(d)},
\end{equation}
where again $n \geq d$. For high powers $k$ we, therefore, save more resources by making use of the parallelized version. In this formula, we assumed $k$ to be constant. We can, however, also consider $k$ that scales with the dimension $d$. For instance, the most resource-demanding scenario is when $k = d/2$. In such a case one obtains
\begin{equation}
    r_{\xgate}(n, d, d/2) \lesssim \frac{3\log_2(n)}{4d}.
\end{equation}
Unless the number of paths exceeds exponentially the dimension, the parallelized scheme of Eq.~\eqref{eq:paral} offers in this scenario substantial savings in resources when compared to the naive approach. For $n < d$ one can perform an analogous analysis.

\section{Losses}
\label{sec:losses}

We can make some rough estimates of the losses of setups presented in the main text by adopting the following simplifications. There are many sources of errors, such as possible beam distortions due to Dove prisms, non-unit conversion efficiency of holograms, different splitting ratios of beam splitters, as well as imperfect reflection of mirrors. Let us assume that all these errors can be modelled as losses quantified by effective mean transmittance $T$ of each optical element where we omit the phase shifters as these can be implemented by a mere path length difference in an interferometer. We also assume that all OAM modes are affected the same way and that $n = d$ in the parallelized scheme.

The number of elements that a photon has to traverse in the universal scheme of Eq.~\eqref{eq:universal} equals approximately $L(d) = d + 10 \log_2(d)$. The transmittance of this scheme thus equals $T^{L(d)}$ and is, therefore, of the same order of magnitude as the transmittance of the scheme of Ref.~\cite{Clements2016} for purely path-encoded unitaries, in which each photon traverses $d$ elements.

The naive implementation of the parallel transformation $U_O^{\mathrm{(par)}}$ with $n = d$ consists of $d$ copies of the universal scheme. When a photon is launched into each of them, there is a chance of $T^{d \, L(d)} = T^{d^2 + 10 d \log_2(d)}$ that all photons make it through the setup and no photon is lost. The parallelized scheme of Eq.~\eqref{eq:paral} has a more complicated structure, where a photon launched into the $j$-th port propagates through a different number $L_j(d)$ of elements. The probability that no photon is lost is given by $T^{L_0(d)} T^{L_1(d)} \ldots T^{L_{d-1}(d)} = T^{\sum_j L_j(d)}$, where the exponent reads $\sum_{j=0}^{d-1} L_j(d) = d^2 + 12 d \log_2(d)$. The simultaneous transmission of $d$ photons through the parallelized scheme is thus quantified by transmittance of $T^{d^2 + 12 d \log_2(d)}$, which differs by a factor of $T^{2 d \log_2(d)}$ from the naive scheme. The per-photon transmittance is thus decreased by a factor of $T^{2 \log_2(d)}$ for the parallelized scheme. If we assume that the effective mean transmittance of each element is $T = 0.9$, this factor does not drop below $0.43$ for dimensions up to $d = 16$.

Let us note that all the schemes for implementing unitaries in OAM presented above share with the purely path-encoded schemes \cite{Reck1994,Clements2016,Guise2018} the fact that the overall transmittance drops down exponentially with the dimension $d$, which is of a particular concern in the real-world experimental implementations.

\section{Periodicity in OAM}
\label{sec:oamper}

The OAM subspace is usually defined as a linear span of eigenstates $\ket{0}_O$, $\ket{1}_O$, $\ket{2}_O$, \ldots, $\ket{d-1}_O$ for a fixed dimension $d$. This subspace can be used in the universal scheme of Eq.~\eqref{eq:universal}. When we want to make use of the periodicity in OAM, the dimension has to be of the form $d = 2^M$ in order that the interferometric implementation of OAM sorters and swap operators works properly for higher-order eigenstates \cite{oamfft2}. For the power-of-two dimensions, the OAM sorter is shown \cite{oamfft} to act like
\begin{equation}
\sorter{d}(\ket{m}_O\ket{0}_P) = \ket{d \cdot \left\lfloor\frac{m}{d} \right\rfloor}_O\ket{\rule{0ex}{2.5ex}m \ \mathrm{mod} \ d}_P.
\end{equation}
This `modulo property' makes sure that OAM eigenstates of the form $\ket{0 + a d}_O$, $\ket{1 + a d}_O$, $\ket{2 + a d}_O$, \ldots, $\ket{d-1 + a d}_O$ for some $a \in \mathbb{Z}$ are not mixed with eigenstates from other OAM subspaces. When an eigenstate $\ket{m + a d}_O$ enters the OAM sorter, it gets transformed into $\ket{a \, d}_O$ propagating along the $m$-th output path. All the aforementioned eigenstates thus leave the OAM sorter along $d$ different paths, but all of them are at that moment equal to $\ket{a \, d}_O$. This way, the path-only implementation $U_P$ then mixes only the terms that correspond to the same subspace, which results in the parallelized operation in the OAM degree of freedom.

From the implementation of the swap operator it follows that whenever the number $n$ of input paths exceeds the number $d$ of output paths, only a specific class of incoming OAM eigenstates gets swapped correctly. Specifically, only eigenstates of the form $\ket{0}_O, \ket{n/d}_O, \ket{2 n/d}_O, \ldots, \ket{k \, n/d}_O$ for $k \in \mathbb{Z}$ can then be used in our parallelized scheme. For these eigenstates, the action of the interferometric implementation of the swap can be summarized as \cite{oamfft2}
\begin{equation}
    \swap_{n, d} \left( \ket{\frac{n}{d} \cdot m}_O\ket{p}_P \right) = \ket{n \cdot \left\lfloor\frac{m}{d} \right\rfloor + p}_O\ket{\rule{0ex}{3.2ex}m \ \mathrm{mod} \ d}_P.
\end{equation}
Since we work only with such $n$ and $d$ that are powers of two, their ratio $n/d$ is an integer.
The above formula is a generalization of the `modulo property' for the swap operator. If $n \leq d$, the action of the swap operator can be written like
\begin{equation}
    \swap_{n, d} \left( \ket{m}_O\ket{p}_P \right) = \ket{\frac{d}{n} \cdot \left( n \cdot \left\lfloor\frac{m}{d} \right\rfloor + p \right)}_O\ket{\rule{0ex}{3.2ex}m \ \mathrm{mod} \ d}_P.
\end{equation}

\bibliographystyle{unsrt}
\bibliography{refs}      

\begin{thebibliography}{10}

\bibitem{erhard_twisted_2018}
Manuel Erhard, Robert Fickler, Mario Krenn, and Anton Zeilinger.
\newblock Twisted photons: new quantum perspectives in high dimensions.
\newblock {\em Light: Science {\&} Applications}, 7(3):17146--17146, March
  2018.

\bibitem{shen_optical_2019}
Yijie Shen, Xuejiao Wang, Zhenwei Xie, Changjun Min, Xing Fu, Qiang Liu, Mali
  Gong, and Xiaocong Yuan.
\newblock Optical vortices 30 years on: {OAM} manipulation from topological
  charge to multiple singularities.
\newblock {\em Light: Science \& Applications}, 8(1):90, October 2019.

\bibitem{Wang2015}
Xi-Lin Wang, Xin-Dong Cai, Zu-En Su, Ming-Cheng Chen, Dian Wu, Li~Li, Nai-Le
  Liu, Chao-Yang Lu, and Jian-Wei Pan.
\newblock Quantum teleportation of multiple degrees of freedom of a single
  photon.
\newblock {\em Nature}, 518(7540):516--519, Feb 2015.

\bibitem{Bouchard2018experimental}
Fr{\'{e}}d{\'{e}}ric Bouchard, Khabat Heshami, Duncan England, Robert Fickler,
  Robert~W. Boyd, Berthold-Georg Englert, Luis~L. S{\'{a}}nchez-Soto, and
  Ebrahim Karimi.
\newblock Experimental investigation of high-dimensional quantum key
  distribution protocols with twisted photons.
\newblock {\em {Quantum}}, 2:111, December 2018.

\bibitem{dada_experimental_2011}
Adetunmise~C. Dada, Jonathan Leach, Gerald~S. Buller, Miles~J. Padgett, and
  Erika Andersson.
\newblock Experimental high-dimensional two-photon entanglement and violations
  of generalized {Bell} inequalities.
\newblock {\em Nature Physics}, 7(9):677--680, September 2011.

\bibitem{PhysRevA.86.012334}
J.~Romero, D.~Giovannini, S.~Franke-Arnold, S.~M. Barnett, and M.~J. Padgett.
\newblock Increasing the dimension in high-dimensional two-photon orbital
  angular momentum entanglement.
\newblock {\em Phys. Rev. A}, 86:012334, Jul 2012.

\bibitem{Fickler13642}
Robert Fickler, Geoff Campbell, Ben Buchler, Ping~Koy Lam, and Anton Zeilinger.
\newblock Quantum entanglement of angular momentum states with quantum numbers
  up to 10,010.
\newblock {\em Proceedings of the National Academy of Sciences},
  113(48):13642--13647, 2016.

\bibitem{Morizur10}
Jean-Fran\c{c}ois Morizur, Lachlan Nicholls, Pu~Jian, Seiji Armstrong, Nicolas
  Treps, Boris Hage, Magnus Hsu, Warwick Bowen, Jiri Janousek, and Hans-A.
  Bachor.
\newblock Programmable unitary spatial mode manipulation.
\newblock {\em J. Opt. Soc. Am. A}, 27(11):2524--2531, Nov 2010.

\bibitem{Labroille2014}
Guillaume Labroille, Bertrand Denolle, Pu~Jian, Philippe Genevaux, Nicolas
  Treps, and Jean-Fran{\c{c}}ois Morizur.
\newblock {Efficient and mode selective spatial mode multiplexer based on
  multi-plane light conversion}.
\newblock {\em Optics Express}, 22(13):15599, jun 2014.

\bibitem{Fontaine2019}
Nicolas~K. Fontaine, Roland Ryf, Haoshuo Chen, David~T. Neilson, Kwangwoong
  Kim, and Joel Carpenter.
\newblock {Laguerre-Gaussian mode sorter}.
\newblock {\em Nature Communications}, 10(1):1865, dec 2019.

\bibitem{Brandt20}
Florian Brandt, Markus Hiekkam\"{a}ki, Fr\'{e}d\'{e}ric Bouchard, Marcus Huber,
  and Robert Fickler.
\newblock High-dimensional quantum gates using full-field spatial modes of
  photons.
\newblock {\em Optica}, 7(2):98--107, Feb 2020.

\bibitem{oamfft2}
Jaroslav Kysela.
\newblock {H}igh-dimensional quantum {F}ourier transform of twisted light.
\newblock {\em Phys. Rev. A}, 104:012413, Jul 2021.

\bibitem{HeisenbergWeylObs}
Ali Asadian, Paul Erker, Marcus Huber, and Claude Kl\"ockl.
\newblock {H}eisenberg-{W}eyl observables: Bloch vectors in phase space.
\newblock {\em Phys. Rev. A}, 94:010301(R), Jul 2016.

\bibitem{Palici_2020}
Alexandra~Maria P{\u{a}}lici, Tudor-Alexandru Isdrail{\u{a}}, Stefan Ataman,
  and Radu Ionicioiu.
\newblock {OAM} tomography with {Heisenberg}{\textendash}{Weyl} observables.
\newblock {\em Quantum Science and Technology}, 5(4):045004, jul 2020.

\bibitem{Garcia-Escartin2011}
Juan~Carlos Garc{\'{i}}a-Escart{\'{i}}n and Pedro Chamorro-Posada.
\newblock {Universal quantum computation with the orbital angular momentum of a
  single photon}.
\newblock {\em Journal of Optics}, 13(6):064022, jun 2011.

\bibitem{Gao2019}
Xiaoqin Gao and Zhengwei Liu.
\newblock Universal quantum computation by a single photon.
\newblock sep 2019.

\bibitem{Reck1994}
Michael Reck, Anton Zeilinger, Herbert~J. Bernstein, and Philip Bertani.
\newblock {Experimental realization of any discrete unitary operator}.
\newblock {\em Physical Review Letters}, 73(1):58--61, jul 1994.

\bibitem{Clements2016}
William~R. Clements, Peter~C. Humphreys, Benjamin~J. Metcalf, W.~Steven
  Kolthammer, and Ian~A. Walmsley.
\newblock Optimal design for universal multiport interferometers.
\newblock {\em Optica}, 3(12):1460--1465, Dec 2016.

\bibitem{Guise2018}
Hubert de~Guise, Olivia Di~Matteo, and Luis~L. S\'anchez-Soto.
\newblock Simple factorization of unitary transformations.
\newblock {\em Phys. Rev. A}, 97:022328, Feb 2018.

\bibitem{Berkhout2010}
Gregorius C.~G. Berkhout, Martin P.~J. Lavery, Johannes Courtial, Marco~W.
  Beijersbergen, and Miles~J. Padgett.
\newblock Efficient sorting of orbital angular momentum states of light.
\newblock {\em Physical Review Letters}, 105(15):153601, oct 2010.

\bibitem{Fickler2017}
Robert Fickler, Manit Ginoya, and Robert~W. Boyd.
\newblock {Custom-tailored spatial mode sorting by controlled random
  scattering}.
\newblock {\em Physical Review B}, 95(16):161108(R), apr 2017.

\bibitem{Leach2002}
Jonathan Leach, Miles~J. Padgett, Stephen~M. Barnett, Sonja Franke-Arnold, and
  Johannes Courtial.
\newblock Measuring the orbital angular momentum of a single photon.
\newblock {\em Physical Review Letters}, 88(25):257901, jun 2002.

\bibitem{Leach2004}
Jonathan Leach, Johannes Courtial, Kenneth Skeldon, Stephen~M. Barnett, Sonja
  Franke-Arnold, and Miles~J. Padgett.
\newblock Interferometric methods to measure orbital and spin, or the total
  angular momentum of a single photon.
\newblock {\em Physical Review Letters}, 92(1):013601, jan 2004.

\bibitem{Garcia-Escartin2008}
Juan~Carlos Garc{\'{i}}a-Escart{\'{i}}n and Pedro Chamorro-Posada.
\newblock {Quantum multiplexing with the orbital angular momentum of light}.
\newblock {\em Physical Review A}, 78(6):062320, dec 2008.

\bibitem{oamfft}
Jaroslav Kysela, Xiaoqin Gao, and Borivoje Daki\ifmmode~\acute{c}\else
  \'{c}\fi{}.
\newblock Fourier transform of the orbital angular momentum of a single photon.
\newblock {\em Phys. Rev. Applied}, 14:034036, Sep 2020.

\bibitem{xgate}
Xiaoqin Gao, Mario Krenn, Jaroslav Kysela, and Anton Zeilinger.
\newblock {Arbitrary d-dimensional Pauli X gates of a flying qudit}.
\newblock {\em Physical Review A}, 99(2):023825, feb 2019.

\bibitem{isdraila_cyclic_2019}
Tudor-Alexandru Isdrail{\u a}, Cristian Kusko, and Radu Ionicioiu.
\newblock Cyclic permutations for qudits in d dimensions.
\newblock {\em Scientific Reports}, 9(1):6337, December 2019.

\bibitem{Note1}
The reason why the path permutations themselves in Fig.~\ref {fig:xgate} are
  not cyclic permutations is that there is an additional permutation that comes
  from the construction of the OAM sorter $S$ and its inverse $S^{-1}$ \cite
  {oamfft}.

\bibitem{zeng_realization_2016}
Qiang Zeng, Tao Li, Xinbing Song, and Xiangdong Zhang.
\newblock Realization of optimized quantum controlled-logic gate based on the
  orbital angular momentum of light.
\newblock {\em Optics Express}, 24(8):8186, April 2016.

\bibitem{kagalwala_single-photon_2017}
Kumel~H. Kagalwala, Giovanni Di~Giuseppe, Ayman~F. Abouraddy, and Bahaa E.~A.
  Saleh.
\newblock Single-photon three-qubit quantum logic using spatial light
  modulators.
\newblock {\em Nature Communications}, 8(1):739, December 2017.

\bibitem{Moreno04}
Ivan Moreno.
\newblock Jones matrix for image-rotation prisms.
\newblock {\em Appl. Opt.}, 43(17):3373--3381, Jun 2004.

\bibitem{padgett_dove_1999}
Miles~J. Padgett and J.~Paul Lesso.
\newblock Dove prisms and polarized light.
\newblock {\em Journal of Modern Optics}, 46(2):175--179, February 1999.

\bibitem{XuBo2005}
XuBo Zou and W.~Mathis.
\newblock Scheme for optical implementation of orbital angular momentum beam
  splitter of a light beam and its application in quantum information
  processing.
\newblock {\em Phys. Rev. A}, 71:042324, Apr 2005.

\bibitem{PhysRevA.94.033847}
Fang-Xiang Wang, Wei Chen, Zhen-Qiang Yin, Shuang Wang, Guang-Can Guo, and
  Zheng-Fu Han.
\newblock Scalable orbital-angular-momentum sorting without destroying photon
  states.
\newblock {\em Phys. Rev. A}, 94:033847, Sep 2016.

\bibitem{Ionicioiu2016}
Radu Ionicioiu.
\newblock {Sorting quantum systems efficiently}.
\newblock {\em Scientific Reports}, 6(1):25356, jul 2016.

\bibitem{erhard_experimental_2018}
Manuel Erhard, Mehul Malik, Mario Krenn, and Anton Zeilinger.
\newblock Experimental {Greenberger}–{Horne}–{Zeilinger} entanglement
  beyond qubits.
\newblock {\em Nature Photonics}, 12(12):759--764, December 2018.

\bibitem{Note2}
In the universal scheme in Sec.~\ref {sec:universal} this was not an issue as
  all the paths contained only the fundamental internal OAM mode.

\bibitem{Zhongeabe8770}
Han-Sen Zhong, Hui Wang, Yu-Hao Deng, Ming-Cheng Chen, Li-Chao Peng, Yi-Han
  Luo, Jian Qin, Dian Wu, Xing Ding, Yi~Hu, Peng Hu, Xiao-Yan Yang, Wei-Jun
  Zhang, Hao Li, Yuxuan Li, et~al.
\newblock Quantum computational advantage using photons.
\newblock {\em Science}, 2020.

\bibitem{Note3}
The parallelized scheme exhibits the same periodicity property with one
  modification. As follows from the construction of the swap operator, when $n
  > d$ the values of input OAM eigenstates have to be chosen like $\protect
  \ensuremath {\left | i \protect \, n/d \right \rangle }$. The formula
  \protect \eqref {eq:oam_paral} is then modified such that $i \to i n /d$ and
  $j \to j n/d$ where we assume that not only the OAM-space dimension but also
  the number of paths is a power of two.

\bibitem{Campbell_12}
Geoff Campbell, Boris Hage, Ben Buchler, and Ping~Koy Lam.
\newblock Generation of high-order optical vortices using directly machined
  spiral phase mirrors.
\newblock {\em Appl. Opt.}, 51(7):873--876, Mar 2012.

\bibitem{Shen_2013}
Yong Shen, Geoff~T Campbell, Boris Hage, Hongxin Zou, Benjamin~C Buchler, and
  Ping~Koy Lam.
\newblock Generation and interferometric analysis of high charge optical
  vortices.
\newblock {\em Journal of Optics}, 15(4):044005, apr 2013.

\end{thebibliography}

\end{document}